# High-Power Mid-IR Few-Cycle Frequency Comb from Quadratic Solitons in an Optical Parametric Oscillator

**Mingchen Liu[1], Robert M. Gray[1], Arkadev Roy[1], Kirk A. Ingold[2], Evgeni Sorokin[3], Irina Sorokina[4], Peter G. Schunemann[5], Alireza Marandi[1], ***

[1]*Department of Electrical Engineering, California Institute of Technology, Pasadena, California, 91125, USA*
[2]*Photonics Research Center, U.S. Military Academy, West Point, New York, 10996, USA*
[3]*Photonics Institute, Vienna University of Technology, 1040 Vienna, Austria*
[4]*Department of Physics, Norwegian University of Science and Technology, N-7491 Trondheim, Norway*
[5]*BAE Systems, P. O. Box 868, MER15-1813, Nashua, New Hampshire, 03061-0868, USA*
[*]*marandi@caltech.edu*

**Abstract:** Powerful and efficient optical frequency combs in the mid-infrared (MIR) spectral region are highly desirable for a broad range of applications. Despite extensive efforts utilizing various techniques, MIR frequency comb sources are still lacking power, efficiency, or bandwidth for many applications. Here, we report the generation of an intrinsically locked frequency comb source centered at 4.18 μm from an optical parametric oscillator (OPO) operating in the simulton regime, in which formation of purely quadratic solitons lead to enhanced performance. We show advantages of operation in the simulton regime in direct experimental comparisons to the conventional regime, which are also supported by simulation and theory. We achieve 565 mW of average power, 900 nm of instantaneous 3-dB bandwidth, 350% slope efficiency, and 44% conversion efficiency, a performance that is superior to previous OPO demonstrations and other sources in this wavelength range. This work opens a new avenue toward MIR frequency comb generation with high power and efficiency and suggests the great potential of soliton generation based on quadratic nonlinearity in the MIR spectral region.

## 1. Introduction

Optical frequency comb generation in the mid-infrared (MIR) spectral region (3-25 μm) has been a subject of intensive research over the past decades, driven by its numerous applications ranging from precise sensing to fundamental science [1], of which notable examples are molecular spectroscopy [2,3], astronomical spectrograph calibration [4,5], and high-harmonic generation [6,7]. Referred to as the "molecular fingerprint region", the MIR portion of the electromagnetic spectrum contains strong rovibrational absorption features of many molecules, the detection of which is useful for a plethora of applications such as medicine, environmental science, agriculture, energy, and defense. In particular, the 3-5 μm band is of high interest as it contains strong absorptions of many important molecules, including greenhouse gases (e.g., carbon dioxide at ~4.2 μm, nitrous oxide at ~4.5 μm and methane at ~3.3 μm), species used in breath analysis (e.g., ethane at ~3.3 μm and carbon monoxide at ~4.7 μm) and major air pollutants (e.g., nitrogen dioxide at ~3.5 μm and sulfur dioxide at ~4 μm) [8,9]. Given its significance, it is highly desirable to produce frequency combs in this band with great power, efficiency, bandwidth, and stability.

The most widely used techniques to produce MIR frequency combs include difference frequency generation (DFG), optical parametric oscillators (OPO), quantum cascade lasers (QCL), microresonators, supercontinuum generation (SCG), and direct MIR lasing. DFG-based sources feature single-pass configuration and passive cancellation of the carrier-envelope offset frequency ($f_{ceo}$) [10–12] but are limited by their relatively low powers and efficiencies. QCLs have been demonstrated to be a promising alternative to optical nonlinear methods for frequency comb generation [13–15] but currently exhibit narrow instantaneous bandwidth and limited spectral coverage. Other MIR frequency comb sources, including microresonators [16], SCG [17,18] and direct MIR lasing [19] are still facing challenges to reach beyond 3.5 μm.

Compared to other techniques, OPOs have high powers and efficiencies with broad spectral coverages and wide tuning ranges [20–25]. Among the various OPO configurations, synchronously pumped degenerate OPOs have been demonstrated to be particularly promising, featuring high conversion efficiencies [26,27], two-octave-wide spectra [28], few-cycle pulses [29], scalability to a multi-GHZ repetition rate [30], and more importantly, intrinsic phase and frequency locking of the output to the pump [31]. However, the demonstrated OPOs with a wavelength coverage beyond 3 μm have either a MIR conversion efficiency smaller than 20% [20,22–24,26] or a limited MIR output power under 250 mW [27,28].

Recently, there has been increasing interest in realization of purely dissipative cavity solitons [32,33] with the promise of frequency comb sources outside the well-developed near-IR region. The temporal simulton, a special form of quadratic solitons characterized by the generation of simultaneous bright-dark solitons of the signal at ω and the pump at 2ω [34,35], has emerged as a novel state of operation in OPOs [36,37]. The simulton-based OPO can be considered as a combination of a degenerate optical parametric amplifier (OPA) and a positively detuned cavity, in which a double balance of energy and timing is achieved [37], as illustrated in **Figure 1**a-d. While the energy balance results from the interplay of dissipation and amplification, the timing balance is rooted in the compensation of cavity detuning by the nonlinear acceleration. Running in an uncommon high-gain low-finesse regime, simulton-based OPOs feature even higher power and efficiency as well as favorable power-dependent bandwidth scaling without losing any advantages of conventional OPOs. Although operation in the simulton regime has offered a promising new avenue for frequency comb generation in the MIR spectral region, it has remained challenging to extend it to longer wavelengths due to an incomplete understanding of its formation requirements and challenges in experiment.

In this article, we demonstrate an OPO working in the simulton regime which generates a frequency comb centered at 4.18 μm with a high average power of 565 mW, a record conversion efficiency of 44%, an instantaneous full-width at half-maximum (FWHM) bandwidth from 3.6 μm to 4.5 μm, and pulses of 45-fs duration, making it an outstanding mid-IR frequency comb source. A direct experimental comparison between the simulton and conventional regimes under the exact same pump condition attributes many of these outstanding characteristics to the simulton formation. Moreover, we perform numerical simulations to capture the behavior exhibited by different regimes of the OPO, which agree well with our experimental results. The simulation also indicates a pathway to further improve the performance of the simulton-based OPO. Lastly, we highlight key features of the simulton build-up dynamics and offer a discussion on the impact of the pump carrier-envelope offset frequency on simulton formation, with many practical implications. This work presents a powerful scheme for MIR frequency comb generation and demonstrates its potential to be extended to longer wavelengths and integrated platforms [38].

## 2. Experimental setup

The experimental setup of the MIR OPO is illustrated in Figure 1e. For the pump, another OPO based on periodically poled lithium niobate with a 250-MHz repetition rate generating pulses centered at 2.09 μm is used. Its average power reaches up to 1290 mW with a FWHM bandwidth around 155 nm. The MIR OPO cavity consists of a bowtie resonator with a tunable cold-cavity time of ~4 ns that can be scanned or locked around the pump repetition period using a piezoelectric actuator. The input coupler ($M_1$) is a flat dielectric-coated mirror that is highly transmissive for the pump range (around 2.09 μm) and highly reflective for the signal range (around 4.18 μm). The focusing and collimating of the beams are provided by two concave gold mirrors ($M_2$ and $M_3$) with a 24-mm radius of curvature, which have a high reflection for both signal and pump. The nonlinear gain is provided by a plane-parallel orientation-patterned gallium phosphide (OP-GaP) crystal with a length of 0.5 mm and a poling period of 92.7 μm for type-0 phase matching at room temperature. The crystal has a broadband anti-reflection coating for both the signal and pump range. The output coupler ($M_4$) is a dielectric mirror coated for broadband transmission (T=25%) from 3.5 μm to 5.5 μm, the value of which is chosen based on a rough experimental optimization of the output coupling using a coated pellicle beamsplitter [27]. The length of the OPO is locked using the "dither-and-lock" procedure described in reference [31].

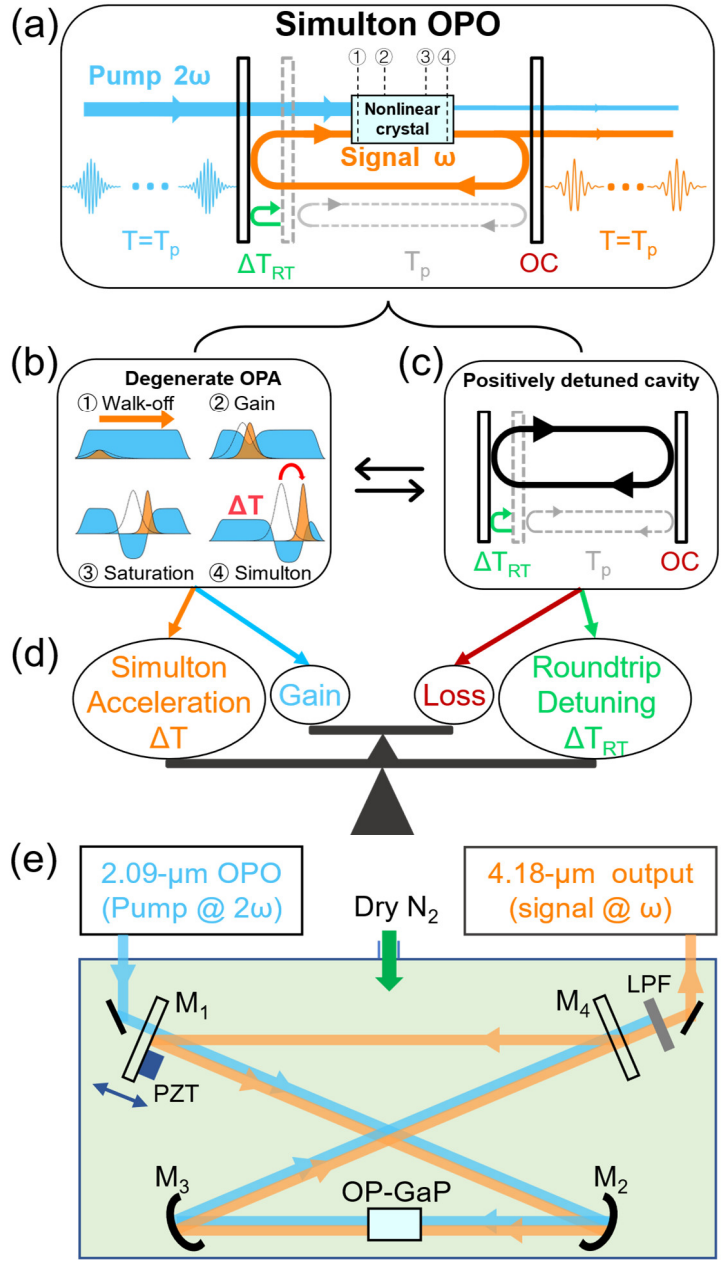

**Figure 1.** The simulton-based OPO. (a) Simplified diagram of the oscillator. $T_p$: Repetition period of the pump pulse and the signal pulse, which is also the cold-cavity round-trip time of the degenerate OPO working in the perfectly synchronous (conventional) regime. $\Delta T_{RT}$: positive timing mismatch between the cavity time of the simulton OPO and $T_p$. OC, partial output coupler. The oscillator can be considered a combination of (b) a degenerate OPA and (c) a positively detuned cavity, striking (d) a double balance of timing and energy. (b) Illustration of simulton formation: signal (orange) at $\omega$ and pump (blue) at $2\omega$. For comparison, uncolored solid lines denote a perfectly synchronous ($\Delta T_{RT}=0$) half-harmonic pulse undergoing linear propagation. ①: on each roundtrip, a small group delay, $\Delta T_{RT}$, is acquired by the resonating signal pulse with respect to the newly in-coupled pump pulse due to the detuning of the cavity roundtrip time. ②③: Passing though the crystal, the signal is amplified by extracting gain from the pump until the pump is depleted, meanwhile accumulating a simulton group advance. ④: Once depleted, the pump forms a dark soliton and co-propagates with the signal at the simulton velocity. (e) Schematic of the 4.18-μm OPO cavity. The cavity length is controlled by mounting $M_1$ on a piezo stage (PZT). The whole cavity resides in a box purged with dry nitrogen to reduce the effects of atmospheric absorption on the OPO operation. Although the OPO can still run without purging, degenerate operation is not possible due to the strong absorption of carbon dioxide centered at 4.2 μm, prohibiting the OPO from operating in the simulton regime. We also contain the measurement instruments for characterization of the OPO output in the purging box to limit the artifacts caused by the atmospheric absorption. LPF, long-pass filer.

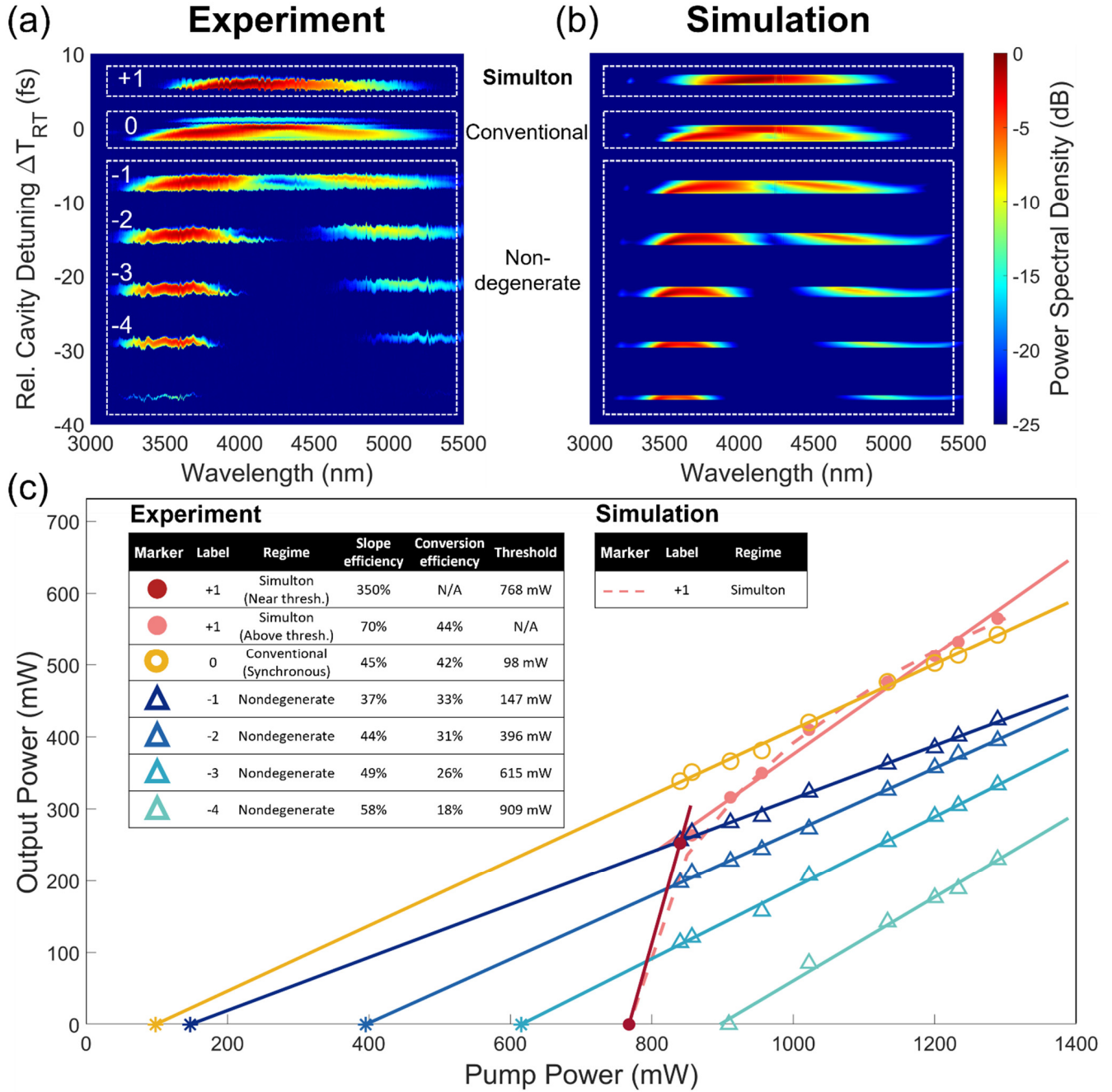

**Figure 2.** (a) Measured signal spectrum as a function of cavity detuning at the highest pump power of 1290 mW, labeled with resonance numbers. The y-axis denotes the relative cavity detuning, the zero of which is set at the center of the conventional resonance. (b) Simulated signal spectrum as a function of cavity detuning corresponding to (a), with the three identified regimes indicated on the left. (c) Output-input power dependencies for each resonance measured with locked cavity lengths. Filled circle, unfilled circles and triangles denote experimental measured points of simulton (+1), conventional (0), and non-degenerate (-1 to -4) resonances, respectively. Solid lines represent their linear fitting for estimation of their slope efficiencies. Note that the thresholds of resonance 0, -1, -2 and -3 cannot be directly measured since the pump is not stable at such low powers; therefore, they are instead estimated by extrapolations of their linearly fitted lines, denoted by asterisks. Two slope efficiencies, one just above the threshold (dark red solid line) and the other well above the threshold (pink solid line), are estimated for the simulton resonance, given its distinct behavior from other resonances. Conversion efficiencies for all resonances are calculated at the highest pump power of 1290 mW. The simulation corresponding to resonance +1 is denoted by the pink dashed curve.

## 3. Results

Because the parametric gain is phase-sensitive, the signal of the OPO only oscillates around the cavity lengths where it acquires a 0 or π phase shift relative to the pump on each roundtrip. This results in signal resonances at a discrete set of cavity lengths, which are separated by about half the signal center wavelength. In the context of this paper, the cavity length is denoted by the deviation of the cold-cavity round-trip time from the repetition period of the pump pulses, i.e. $\Delta T_{RT}$. Depending on the $\Delta T_{RT}$, the different resonances can be classified into three regimes of operation: simulton ($\Delta T_{RT} > 0$), conventional ($\Delta T_{RT} \cong 0$, also referred to as "synchronous"), and nondegenerate ($\Delta T_{RT} < 0$) [36,37]. Note that the positive $\Delta T_{RT}$ corresponds to a longer cavity length. **Figure 2**a shows the measured output spectrum as a function of relative cavity timing detuning ($\Delta T_{RT}$) at the highest pump power of 1290 mW. When the cold cavity is most nearly synchronized to the pump repetition period ($\Delta T_{RT} \cong 0$), the OPO is identified to run in the conventional (synchronous) regime (labeled "0"), which has a degenerate spectrum and the lowest threshold. One additional degenerate resonance, the simulton regime (labeled "+1"), is found when the cavity is positively detuned. Conversely, when the cavity is negatively detuned, the OPO operates in the nondegenerate regime, with the output spectra split into distinguishable signal and idler bands (labeled "-1, -2, -3, -4"). With parameters comparable to the experiment, a simulation of the output spectrum as a function of $\Delta T_{RT}$ is conducted (see Supporting Information for additional details), which is depicted in Figure 2b. The simulation exhibits a good agreement with the experimental result for all three regimes.

Figure 2c present output-input power dependencies for each resonance measured with locked cavity lengths. With the lowest threshold, the conventional regime has a slope efficiency of 45% and a conversion efficiency of 42%. For the nondegenerate resonances, as $\Delta T_{RT}$ becomes increasingly negative, the thresholds increase uniformly, with conversion efficiencies decreasing and limited to less than 35%. In contrast, the simulton resonance has an irregularly located high threshold and the highest conversion efficiency of 44%. Furthermore, it measures a ~350% slope efficiency near the threshold and a ~70% slope efficiency well above the threshold, much higher than those of the conventional and nondegenerate regimes.

To demonstrate the difference between the power-dependent bandwidth scaling of the conventional and simulton regimes, we measure spectra of the signal at each output power corresponding to the experimental points in Figure 2c, the results of which are shown in **Figure 3**a-c. In the simulton regime, as the power increases, the bandwidth of the signal spectrum increases if the pump power is not too high (Figure 3a), while in the conventional regime, it monotonically decreases (Figure 3b). These tendencies are in accordance with the simulton theory [37] and conventional box-pulse scaling [36]. It should be noted that at the three highest pump powers, the signal bandwidth of the simulton regime stops broadening further, which also agrees with our theoretical prediction that the simulton theory would fail if the signal is too far above threshold [36]. This transition from simulton scaling to box-pulse scaling far above threshold also accounts for the observed nonlinear reduction in the simulton slope efficiency in Figure 2c. Nonetheless, at high pump power around 1200 mW, the simulton regime wins about 40% in bandwidth. This power-dependent signal spectral characterization shows that the simulton regime outperforms the conventional regime not only in power and efficiency but also in spectral bandwidth. At the highest available pump power of 1290 mW, the FWHM bandwidth of the signal spectra for simulton and conventional regimes are 14 THz and 10 THz, which can support pulses as short as 22 fs and 32 fs, respectively. Figure 3d depicts the interferometric autocorrelation of the simulton pulse measured by a two-photon extended-InGaAs detector, together with its fitted pulse intensity. This measurement corresponds to a FWHM pulse width of ~45 fs, assuming no chirp. However, chirp exists due to the dispersion from the substrates of the output coupler (1-mm ZnSe) and two long pass filters (1-mm Ge and 1-mm Si) in the path to the autocorrelator.

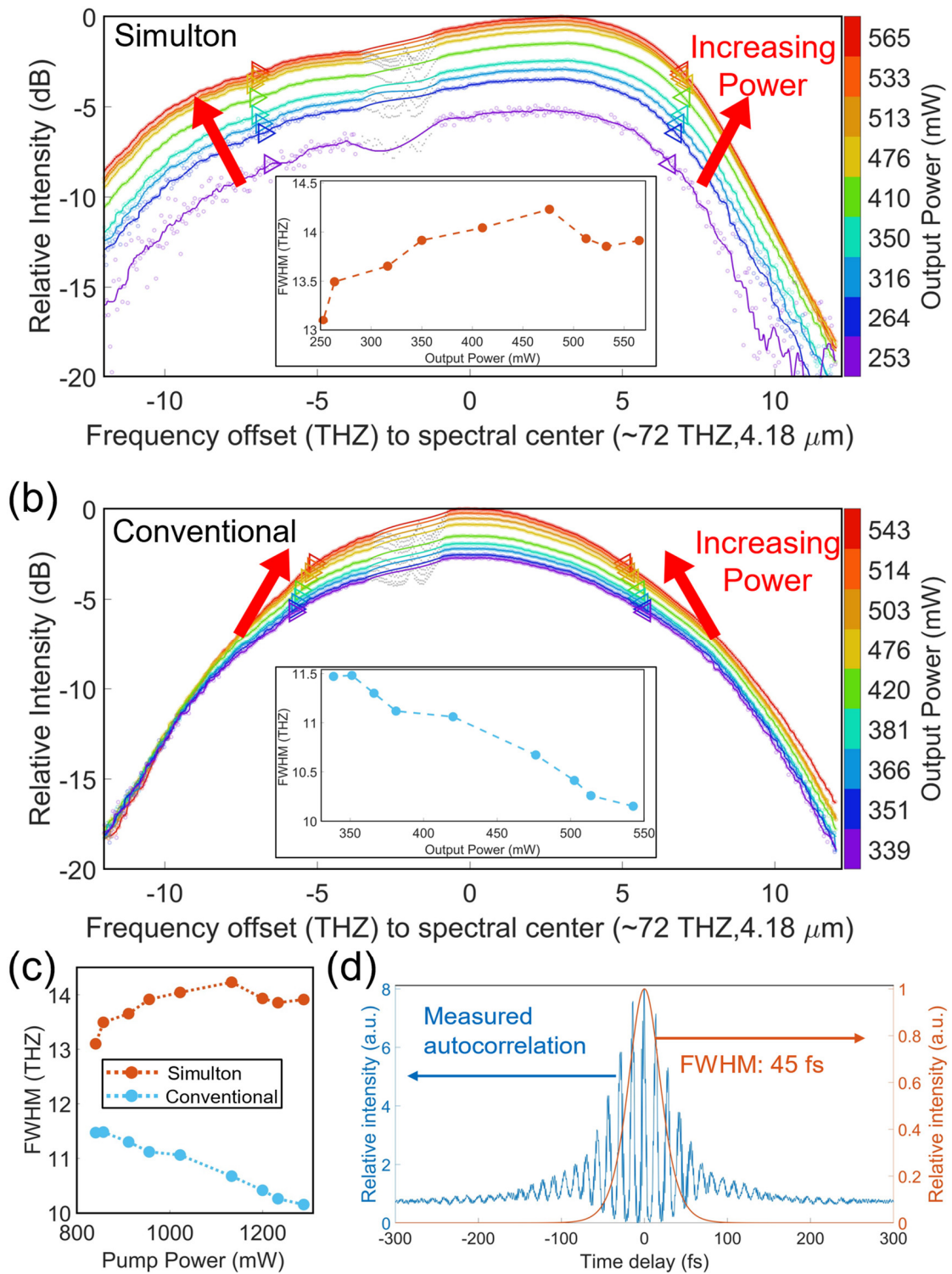

**Figure 3.** (a)(b) Spectra recorded as a function of output power for the OPO working in the (a) simulton regime and (b) conventional (synchronous) regime. The signal power and corresponding FWHM bandwidth for each curve are presented in the insets. The FWHM bandwidths are also denoted by the triangular arrows on the curves. The unfilled circles denote the raw data points obtained by the Fourier-transform infrared spectroscopy (FTIR), and curves present the interpolation of them for a better visualization of results and estimation of FWHM bandwidths. Note that a portion of the raw data near the strong atmospheric absorption around 4.2 μm is discarded during the interpolation, which is denoted by smaller filled grey circles. (c) FWHM bandwidths of the signal spectra as a function of pump power for both regimes, corresponding to (a) and (b). Solid circles denote experimentally measured points and dotted curves are to guide the eye. (d) Two-photon interferometric autocorrelation (blue) and fitted intensity (red) of the signal pulse at the highest pump power of 1290 mW, for the OPO working in the simulton regime.

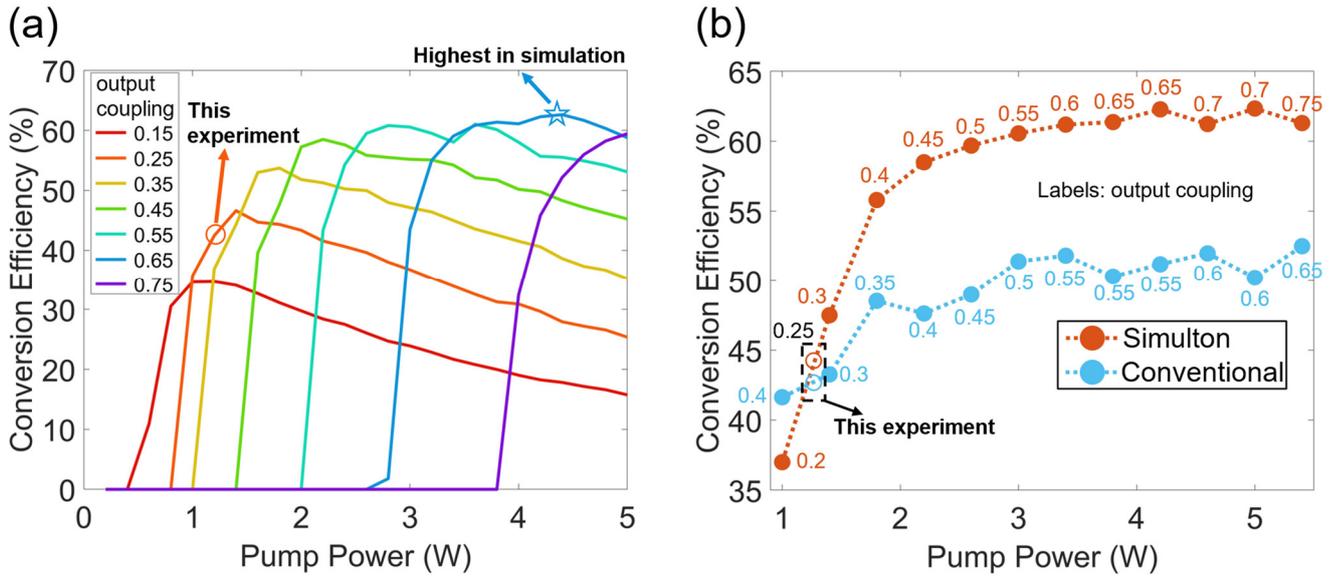

**Figure 4.** Numerical simulation of the simulton OPO with different output coupling ratio. (a) Conversion efficiency (in percentage) as a function of pump power. Curves in different colors denote different output coupling ratio (in decimal), as indicated in the legend box. The orange circle denotes the result that is realized in our experiment, and the blue pentagram denotes the suggested highest conversion efficiency that can be achieved by this simulton-OPO. (b) Highest conversion efficiency (in percentage) that the OPO can achieve under different pump power. The red circles denote the results (in percentage) in the simulton regime, labeled with the output coupling ratio (red decimal) that should be used. The blue circles, as a comparison to the red ones, denote the results in the conventional regime, also labeled with the corresponding output coupling ratio (blue decimal). The unfilled red and blue circles denote our experimental results of simulton and conventional regime, respectively.

In this experiment, the efficiency advantage of the simulton regime is limited by the available pump power. To further demonstrate the efficiency potential of the simulton regime, we use numerical simulations with higher pump power under different output coupling ratios, and the results are shown in **Figure 4**a. As suggested by the simulation, higher output coupling should be employed with higher pump power to realize higher conversion efficiency, and the conversion efficiency can be improved to as high as 63% if 4.2-W pump power and 0.65(65%) output coupling ratio are used. This simulation result can also be used as a design rule for choosing the output coupling of the simulton OPO under different available pump powers. Figure 4b presents the highest possible conversion efficiency that the simulton regime can reach under different pump powers (red circles), with the labels of the corresponding output coupling (red decimals). As a comparison, under each pump power, the highest conversion efficiencies that the conventional regime can achieve are also plotted in Figure 4b (blue circles), labeled with the corresponding output coupling ratios (blue decimals). The comparison shows that the return of the simulton regime can increase sharply with increasing pump power. It is worth noting that the pump intensity used in the simulation is similar to those of the previously demonstrated experiments [28,39], which is expected to be below the damage threshold of OP-GaP.

## 4. Discussion

As is evident from our experimental and theoretical results, developing a better understanding of simulton formation is crucial for further improvement of the OPO performance. Here, we use our simulation results to build upon the understanding of simulton formation dynamics presented in [37] and offer practical tools for optimizing simulton performance. We begin with a discussion of the relationship between the simulton formation dynamics and the high slope efficiencies and conversion efficiencies offered by this regime. **Figure 5**a shows schematically the interaction between pump (blue) and signal (orange) in the single-pass OPA process which occurs each roundtrip in the OPO. From the crystal input (left) to its output (right), the signal walks through the pump, depleting it and extracting gain in the process. Two terms contribute to the walk-off, the product of the group velocity mismatch (GVM), *u*, with the crystal length, *L*, and the nonlinear timing advance due to simulton acceleration, $\Delta T$, leading to a combined walk-off of $uL + \Delta T$ relative to the signal starting position. This walk-off along the fast time axis and simultaneous depletion of the pump by the signal determines the available gain for the signal. For the signal to resonate above threshold, the gain extracted in consecutive roundtrips must consistently overcome the loss.

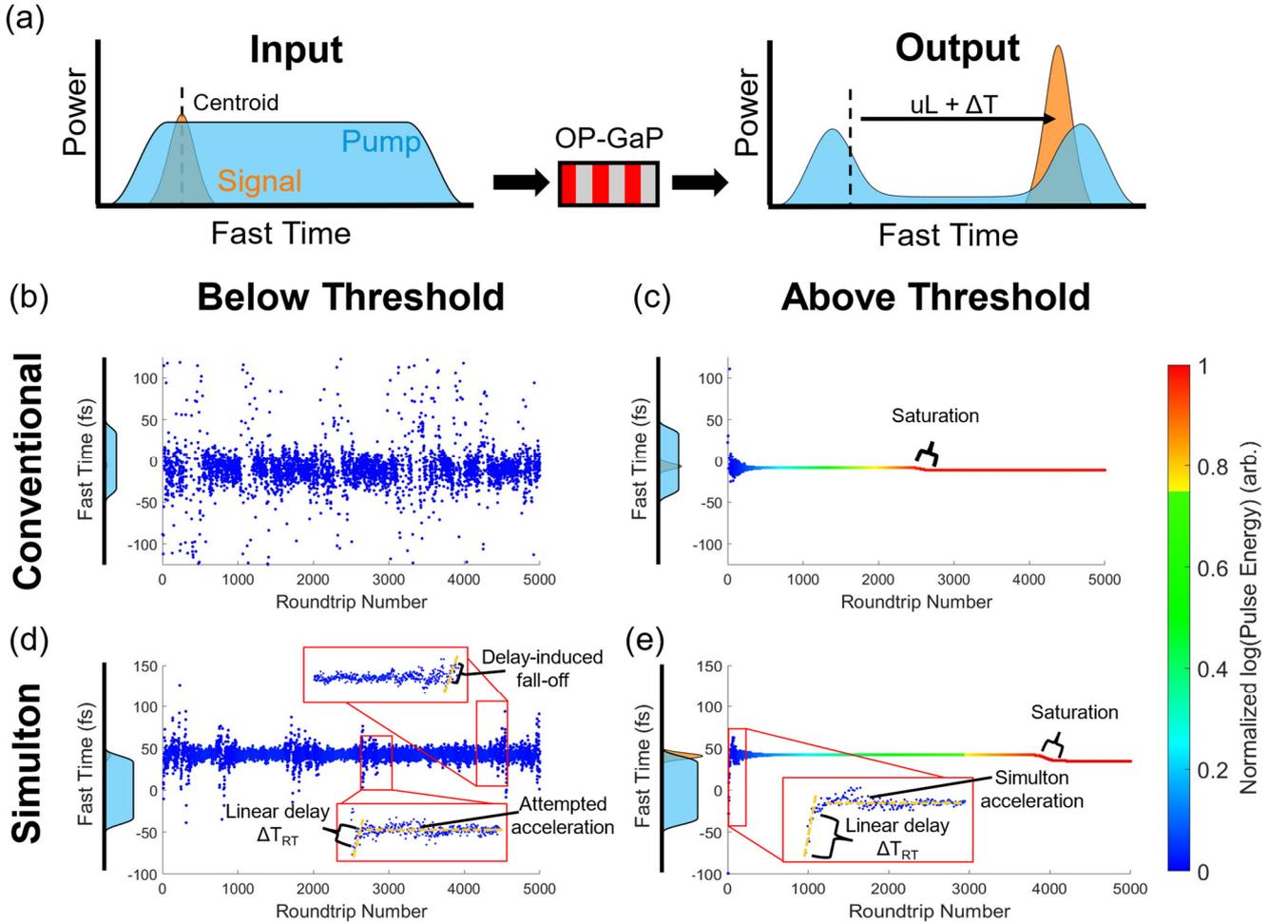

**Figure 5.** Depiction of simulton formation dynamics as compared to the conventional regime. (a) Schematic illustration of the pump-signal interaction in the crystal. Between the input (left) and output (right) facets of the crystal, the signal (orange) walks through the pump (blue) from its starting point according to product of the GVM, $u$, with the crystal length, $L$, plus any timing advance, $\Delta T$, due to simulton acceleration. This walk-off and concurrent pump depletion define a gain window for the signal along the fast time axis. (b-e) Simulated comparison of simulton (b, c) and conventional (d, e) signal centroid positions along the fast time axis as a function of roundtrip number in the below (b, d) and above (c, e) threshold cases. The normalized logarithm of the pulse energy is indicated by the color of each point. (b) The conventional case below threshold has few discernible features as noise and loss dominate; noise amplified by the pump is unable to build up. (c) Above threshold, the noise quickly builds into a strong signal pulse. A small timing shift is observed when the gain saturates. (d) For the simulton, noise amplified by the pump experiences a linear delay due to the cavity detuning $\Delta T_{RT}$. Below threshold, since the gain cannot enable sufficient acceleration for the timing condition to be satisfied, the delay results in the signal falling out of the gain window before signal build up can be achieved. (e) The simulton goes above threshold when the gain is sufficient for the simulton acceleration to compensate the linear delay and enable signal build up. As in the conventional case, a small timing shift is observed when the gain saturates.

Figure 5b-e shows the signal centroid position at the input facet of the crystal as a function of roundtrip, with the color indicating the normalized logarithm of the pulse energy, to illustrate the signal build up dynamics in the conventional and simulton regimes. In the conventional case (Figure 5b-c), there is no timing condition, so the signal goes above threshold as soon as the gain experienced through the linear walk-off, $uL$, equals the loss. This results in there being no clear trends in the dynamics of the signal below threshold, as shown in Figure 5b, with amplified noise being unable to build up. Above threshold,

the noise quickly grows into a strong signal pulse, located near the center of the gain window set by the pump (Figure 5c). In the simulton case (Figure 5d-e), however, the cavity detuning, $\Delta T_{RT}$, creates a linear delay which causes the amplified signal noise to move away from the gain window set by the pump. Since the simulton acceleration relies on pump depletion by the leading edge of the signal and subsequent back-conversion of the trailing edge, going above threshold requires that the simulton build up enough to accelerate and satisfy the timing condition before falling out of the gain window due to the linear delay. This intertwines the simulton energy and timing conditions, as large simulton acceleration requires the presence of high enough signal gain. Figure 5d shows the case of the below threshold simulton, in which the delayed and amplified noise attempts to build up but ultimately does not experience enough gain to sufficiently accelerate and falls off. By contrast, Figure 5e shows the simulton build-up above threshold, in which the simulton acceleration leads to a timing advance, $\Delta T$, which compensates the delay, $\Delta T_{RT}$. Unlike the conventional case, the simulton just above threshold builds up near the edge of the gain window. These observations correspond to important features of the simulton. First, the high threshold for the simulton is a consequence of the requirement for there to be high gain for the timing condition to be satisfied. The burst slope efficiency of the simulton near threshold then results as the acceleration pulls the simulton to the center of the gain window. Once well-confined to the gain window, the longer walk-off for the simulton due to the additional nonlinear acceleration term, $\Delta T$, suggests that simulton operation can often enable more efficient extraction of the pump gain. This gives rise to observed trends of slope efficiencies and overall conversion efficiencies for simulton operation that exceed what is observed for conventional OPO operation. Practically, one should match $uL + \Delta T$ to the length of the gain window defined by the pump for optimum signal generation.

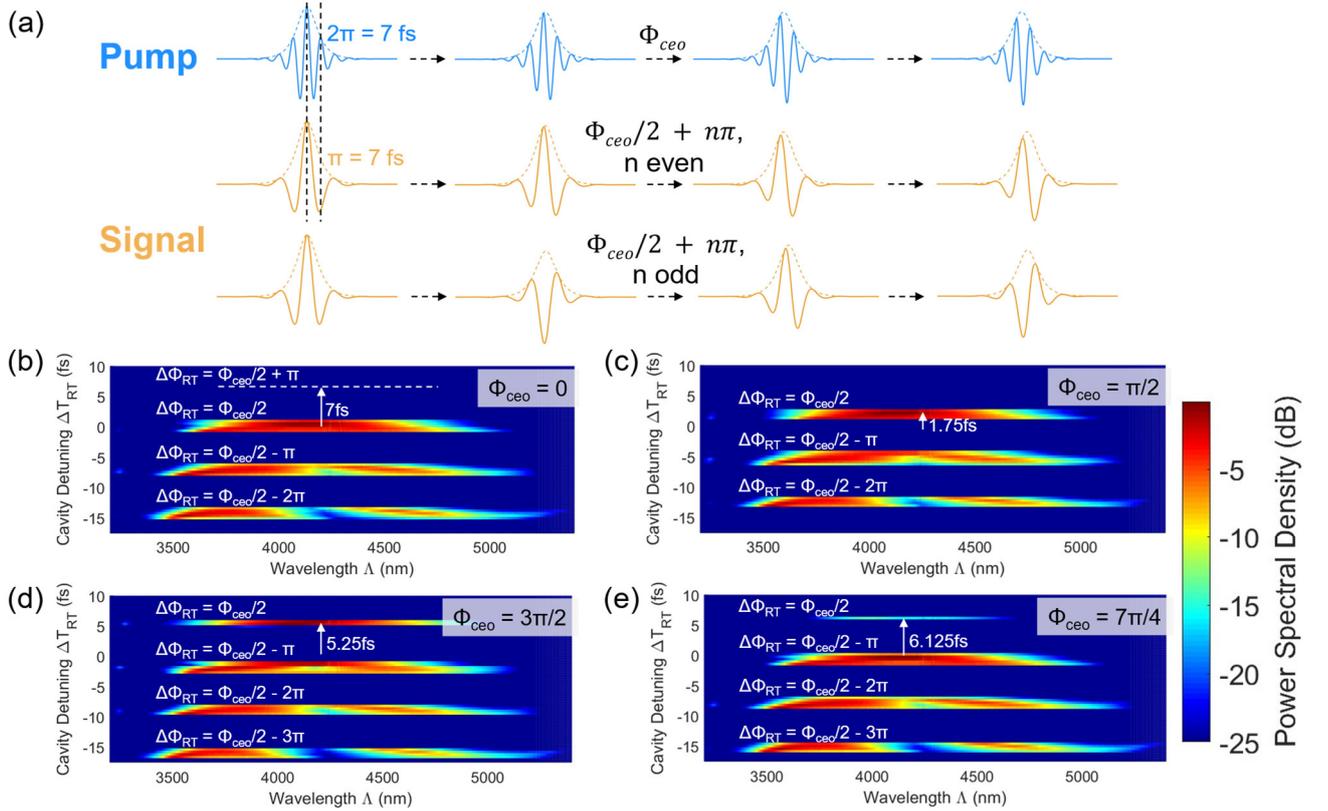

**Figure 6.** Simulated impact of the carrier-envelope offset phase, $\Phi_{ceo}$, on simulton performance. (a) Illustration of the pump-signal phase relationships for the OPO resonances. For a pulse-to-pulse phase slip of $\Phi_{ceo}$ in the pump, the OPO can resonate when the roundtrip phase accumulation, $\Delta\Phi_{RT}$, is $\Phi_{ceo}/2 + n\pi$, where $n$ is an integer, with $\pi$ signal phase corresponding to a detuning of 7 fs between resonances. The upper signal branch depicts the case of n even, where a pulse-to-pulse phase of $\Phi_{ceo}/2$ is accrued, while the lower shows n odd, which adds an additional $\pi$ phase shift between consecutive pulses. (b) First three cavity resonances for $\Phi_{ceo} = 0$, showing that no simulton behavior is observed, with the topmost resonance occurring at $\Delta T_{RT} = 0$ and exhibiting both strongly degenerate and nondegenerate features. (c) As $\Phi_{ceo}$ is increased to $\pi/2$, simulton-like behavior is observed as the topmost resonance becomes more positively detuned. (d) Example of a strongly degenerate simulton at $\Phi_{ceo} = 3\pi/2$, showing that tuning of $\Phi_{ceo}$ to this region can enable optimum

simulton performance for the given gain. (e) As $\Phi_{ceo}$ is further increased beyond $7\pi/4$, the simulton disappears as the timing condition can no longer be satisfied by the gain, which is why no simulton resonance is observed in panel (b) at the $\Phi_{ceo}/2 + \pi$ resonance, indicated by the white, dashed line.

An additional parameter which is critical to simulton formation is the carrier-envelope offset frequency ($f_{ceo}$) of the pump, the impact of which is illustrated in **Figure 6**. The pump $f_{ceo}$ imposes a phase $\Phi_{ceo}$ between consecutive pump pulses. Due to the aforementioned phase-sensitive gain which demands a relative phase of 0 or $\pi$ between pump and signal for signal build-up, this pulse-to-pulse phase shift in the pump must be mirrored in the signal, as shown in Figure 6a. For signal resonance to occur, the phase accumulated in the roundtrip must be $\Phi_{ceo}/2 + n\pi$, where $n$ is an integer, with $\pi$ signal phase corresponding to a detuning of 7 fs between resonances for the 4.18 μm OPO. The case for $n$ even is shown by the upper signal branch while the case where $n$ is odd, in which the signal accumulates an additional phase shift of $\pi$ between pulses, is depicted in the lower signal branch. The signal roundtrip phase accumulation, $\Phi(\Omega)$, relative to a perfectly synchronous signal pulse is given by $\Phi(\Omega) = \Delta T_{RT}(\pi c/\lambda_{2\omega} + \Omega) + \Delta\Phi(\Omega)$, where $\Omega$ is the normalized frequency, $\Delta T_{RT}$ is the detuning as defined previously, $c$ is the speed of light, $\lambda_{2\omega}$ is the pump wavelength, and $\Delta\Phi(\Omega)$ represents the higher-order effects of dispersion from mirrors and additional cavity elements [37]. From this equation, we see the required pulse-to-pulse phase shift for the signal can be achieved through varying $\Delta T_{RT}$ such that the constant phase term $\Delta\Phi_{RT} = \Delta T_{RT}\pi c/\lambda_{2\omega} = \Phi_{ceo}/2 + \pi n$, the desired phase shift. In other words, $\Phi_{ceo}$ determines the detuning values where the OPO can resonate. However, varying $\Delta T_{RT}$ also causes a change in the signal timing, modeled through the linear phase term $\Delta T_{RT}\Omega$, and consequently the simulton threshold and slope efficiency. This implies that, through tuning of the pump $f_{ceo}$, one can adjust the timing of the signal resonances to optimize simulton performance.

Figure 6b-e show the simulated signal resonances for a few values of $\Phi_{ceo}$. As the focus of this study is on simulton behavior, only the first few resonances are shown. Resonance labels denote the roundtrip phase, $\Delta\Phi_{RT}$, acquired by the signal, and an additional label denotes the timing shift from 0 fs of the most positively detuned resonance with $\Delta\Phi_{RT} = \Phi_{ceo}/2$. As seen in Figure 6b, when $\Phi_{ceo} = 0$, the signal at $\Delta\Phi_{RT} = \Phi_{ceo}/2$ is a typical conventional resonance with $\Delta T_{RT} = 0$. Note that no simulton exists at the expected location, $\Delta\Phi_{RT} = \Phi_{ceo}/2 + \pi$ (shown by the white, dashed line), due to a lack of gain. For values between 0 and $\pi$, the $\Delta\Phi_{RT} = \Phi_{ceo}/2$ resonance becomes simulton-like but still behaves more like a conventional OPO, with both non-degenerate and degenerate regions as shown in Figure 6c. Beyond $\Phi_{ceo} = \pi$, a strong simulton is observed, like that shown in Figure 6d for $\Phi_{ceo} = 3\pi/2$. In tandem with the $\Delta\Phi_{RT} = \Phi_{ceo}/2$ resonance transitioning from more conventional behavior to the simulton regime, we also observe strengthening of the degenerate side of the $\Delta\Phi_{RT} = \Phi_{ceo}/2 - \pi$ resonance such that it behaves more like a conventional OPO as it approaches $\Delta T_{RT} = 0$. Finally, as exemplified by Figure 6e, when $\Phi_{ceo}$ is greater than $1.5\pi$, the power in the resonance quickly drops off as the gain struggles to satisfy the timing condition, with the OPO repeating the resonance structure shown in Figure 6b as $\Phi_{ceo}$ approaches $2\pi$. Our experiment corroborates these findings, as we observe such a shift in simulton behavior as the $\Phi_{ceo}$ is varied, with presented data representing the strongest observed simulton. This suggests that, for given experimental values of gain and loss, $\Phi_{ceo}$ can be a crucial experimental parameter for optimizing simulton performance.

## 5. Conclusions

In summary, we present the generation of a high-power and efficient MIR frequency comb centered at 4.18 μm based on an OPO operating in the simulton regime, achieving 565-mW average power together with a 14-THz instantaneous FWHM bandwidth, sub-three-cycle pulses, a 350% slope efficiency near the threshold, a 70% slope efficiency above the threshold, and a record high 44% conversion efficiency. By a direct comparison with the conventional regime in terms of cavity detuning, output power, threshold, slope and conversion efficiency, instantaneous bandwidth, and bandwidth scaling, we are able to ascribe the favorable performances of this novel regime to the simulton formation, based on the agreement between the experiment, simulation, and theory. The performance of this simulton OPO is expected to be further improved by tuning the pump and optimizing dispersion and loss within the cavity. Especially, the numerical simulation suggests that a conversion efficiency >60% can be achieved if 4-W pump power and 65% output coupling are employed. Finally, we further explore simulton build-up dynamics and analyze the impact of the pump $f_{ceo}$ on simulton formation, and we tie these results to practical design considerations for a simulton OPO. In addition, we want to emphasize that the half-harmonic signal of femtosecond degenerate OPOs are frequency combs that are intrinsically phase- and frequency- locked to their pump combs, which is well-established in previous works [26,31,40]. Very recently, we have used the OPO presented in this work as the comb source for a dual-comb-based spectroscopy experiment in the mid-IR [41], enabled by its comb character and intrinsic locking. Moreover, it is experimentally shown that the CEO frequency noise can be reduced by 6 dB through the half-harmonic generation [42].

In short, this work paves the way to realization of a compelling new source of ultrashort-pulse frequency combs in the mid-infrared region which can benefit numerous applications, for example, spectroscopy methods that require high-power, broad-band and short-pulse MIR frequency combs [40,41]. This work sheds new lights on soliton generation based on the quadratic nonlinearity, and its potential in the MIR region. Recent advances in integrated quadratic platforms [33,38,43] promise on-chip realization of such sources in the future.

**Acknowledgments.**
M.L. and R.M.G. contributed equally to this work. The authors gratefully acknowledge support from National Aeronautics and Space Administration (NASA)/ Jet Propulsion Laboratory (JPL). This article was funded by Air Force Office of Scientific Research (AFOSR) award FA9550-20-1-0040 and National Science Foundation (NSF) Grant No. 1846273.

**Conflict of Interest.**
The authors declare no conflicts of interest.

**Data Availability.**
Data underlying the results presented in this paper are not publicly available at this time but may be obtained from the authors upon reasonable request.

**Keywords**
optical parametric oscillator, quadratic soliton, mid-infrared, frequency comb

**REFERENCES**

1. A. Schliesser, N. Picqué, and T. W. Hänsch, Nature Photonics **6**, 440 (2012).
2. N. Picqué and T. W. Hänsch, Nature Photonics **13**, 146 (2019).
3. F. Keilmann, C. Gohle, and R. Holzwarth, Optics Letters **29**, 1542 (2004).
4. T. Steinmetz, T. Wilken, C. Araujo-Hauck, R. Holzwarth, T. W. Hänsch, L. Pasquini, A. Manescau, S. D'Odorico, M. T. Murphy, T. Kentischer, W. Schmidt, and T. Udem, Science **321**, 1335 (2008).
5. J. Bland-Hawthorn and P. Kern, Opt. Express **17**, 1880 (2009).
6. T. Popmintchev, M.-C. Chen, D. Popmintchev, P. Arpin, S. Brown, S. Ališauskas, G. Andriukaitis, T. Balčiunas, O. D. Mücke, A. Pugzlys, A. Baltuška, B. Shim, S. E. Schrauth, A. Gaeta, C. Hernández-García, L. Plaja, A. Becker, A. Jaron-Becker, M. M. Murnane, and H. C. Kapteyn, Science **336**, 1287 (2012).
7. S. Ghimire and D. A. Reis, Nature Physics **15**, 10 (2019).
8. M. Vainio and L. Halonen, Phys. Chem. Chem. Phys. **18**, 4266 (2016).
9. B. Henderson, A. Khodabakhsh, M. Metsälä, I. Ventrillard, F. M. Schmidt, D. Romanini, G. A. D. Ritchie, S. te Lintel Hekkert, R. Briot, T. Risby, N. Marczin, F. J. M. Harren, and S. M. Cristescu, Appl. Phys. B **124**, 161 (2018).
10. I. Pupeza, D. Sánchez, J. Zhang, N. Lilienfein, M. Seidel, N. Karpowicz, T. Paasch-Colberg, I. Znakovskaya, M. Pescher, W. Schweinberger, V. Pervak, E. Fill, O. Pronin, Z. Wei, F. Krausz, A. Apolonski, and J. Biegert, Nature Photonics **9**, 721 (2015).
11. G. Soboń, T. Martynkien, P. Mergo, L. Rutkowski, and A. Foltynowicz, Opt. Lett. **42**, 1748 (2017).
12. A. J. Lind, A. Kowligy, H. Timmers, F. C. Cruz, N. Nader, M. C. Silfies, T. K. Allison, and S. A. Diddams, Phys. Rev. Lett. **124**, 133904 (2020).

13. A. Hugi, G. Villares, S. Blaser, H. C. Liu, and J. Faist, Nature **492**, 229 (2012).
14. J. Faist, G. Villares, G. Scalari, M. Rösch, C. Bonzon, A. Hugi, and M. Beck, Nanophotonics **5**, 272 (2016).
15. M. Razeghi, Q. Lu, D. Wu, and S. Slivken, in *Terahertz Emitters, Receivers, and Applications IX* (SPIE, 2018), **10756**, pp. 21–30.
16. M. Yu, Y. Okawachi, A. G. Griffith, N. Picqué, M. Lipson, and A. L. Gaeta, Nat Commun **9**, 1869 (2018).
17. D. Grassani, E. Tagkoudi, H. Guo, C. Herkommer, F. Yang, T. J. Kippenberg, and C.-S. Brès, Nature Communications **10**, 1553 (2019).
18. H. Guo, H. Guo, W. Weng, J. Liu, F. Yang, W. Hänsel, C. S. Brès, L. Thévenaz, R. Holzwarth, and T. J. Kippenberg, Optica **7**, 1181 (2020).
19. S. Vasilyev, V. Smolski, J. Peppers, I. Moskalev, M. Mirov, Y. Barnakov, S. Mirov, S. Mirov, and V. Gapontsev, Opt. Express **27**, 35079 (2019).
20. F. Adler, K. C. Cossel, M. J. Thorpe, I. Hartl, M. E. Fermann, and J. Ye, Opt. Lett. **34**, 1330 (2009).
21. Y. Kobayashi, K. Torizuka, A. Marandi, R. L. Byer, R. A. McCracken, Z. Zhang, and D. T. Reid, J. Opt. **17**, 094010 (2015).
22. L. Maidment, P. G. Schunemann, and D. T. Reid, Opt. Lett. **41**, 4261 (2016).
23. K. Iwakuni, G. Porat, T. Q. Bui, B. J. Bjork, S. B. Schoun, O. H. Heckl, M. E. Fermann, and J. Ye, Appl. Phys. B **124**, 128 (2018).
24. C. F. O'Donnell, S. C. Kumar, P. G. Schunemann, and M. Ebrahim-Zadeh, Opt. Lett. **44**, 4570 (2019).
25. C. Ning, P. Liu, Y. Qin, and Z. Zhang, Opt. Lett. **45**, 2551 (2020).
26. A. Marandi, K. A. Ingold, M. Jankowski, and R. L. Byer, Optica **3**, 324 (2016).
27. E. Sorokin, A. Marandi, P. G. Schunemann, M. M. Fejer, R. L. Byer, and I. T. Sorokina, Opt. Express **26**, 9963 (2018).
28. Q. Ru, T. Kawamori, P. G. Schunemann, S. Vasilyev, S. B. Mirov, S. B. Mirov, and K. L. Vodopyanov, Opt. Lett. **46**, 709 (2021).
29. R. A. McCracken and D. T. Reid, Opt. Lett. **40**, 4102 (2015).
30. R. A. McCracken, Y. S. Cheng, and D. T. Reid, in *Conference on Lasers and Electro-Optics* (2018), p. FTh1M.1.
31. A. Marandi, N. C. Leindecker, V. Pervak, R. L. Byer, and K. L. Vodopyanov, Opt. Express **20**, 7255 (2012).
32. A. Roy, R. Nehra, S. Jahani, L. Ledezma, C. Langrock, M. Fejer, and A. Marandi, Nat. Photon. **16**, 162 (2022).
33. A. W. Bruch, X. Liu, Z. Gong, J. B. Surya, M. Li, C.-L. Zou, and H. X. Tang, Nature Photonics **15**, 21 (2021).
34. S. Akhmanov, A. Chirkin, K. Drabovich, A. Kovrigin, R. Khokhlov, and A. Sukhorukov, IEEE Journal of Quantum Electronics **4**, 598 (1968).
35. S. Trillo, Opt. Lett. **21**, 1111 (1996).
36. R. Hamerly, A. Marandi, M. Jankowski, M. M. Fejer, Y. Yamamoto, and H. Mabuchi, Phys. Rev. A **94**, 063809 (2016).
37. M. Jankowski, A. Marandi, C. R. Phillips, R. Hamerly, K. A. Ingold, R. L. Byer, and M. M. Fejer, Phys. Rev. Lett. **120**, 053904 (2018).
38. L. Ledezma, L. Ledezma, R. Sekine, Q. Guo, R. Nehra, S. Jahani, and A. Marandi, Optica **9**, 303 (2022).
39. L. Maidment, O. Kara, P. G. Schunemann, J. Piper, K. McEwan, and D. T. Reid, Appl. Phys. B **124**, 143 (2018).
40. A. V. Muraviev, V. O. Smolski, Z. E. Loparo, and K. L. Vodopyanov, Nature Photonics **12**, 209 (2018).
41. M. Liu, R. M. Gray, A. Roy, C. R. Markus, and A. Marandi, arXiv:2107.08333 [physics] (2021).
42. C. Wan, P. Li, A. Ruehl, and I. Hartl, Opt. Lett., OL **43**, 1059 (2018).
43. M. Jankowski, M. Jankowski, M. Jankowski, N. Jornod, N. Jornod, C. Langrock, B. Desiatov, A. Marandi, M. Lončar, and M. M. Fejer, Optica **9**, 273 (2022).

Supplementary Materials for

# High-Power Mid-IR Few-Cycle Frequency Comb from Quadratic Solitons in an Optical Parametric Oscillator

**Mingchen Liu[1, †], Robert M. Gray[1, †], Arkadev Roy[1], Kirk A. Ingold[2], Evgeni Sorokin[3], Irina Sorokina[4], Peter G. Schunemann[5], Alireza Marandi[1, *]**

[1]*Department of Electrical Engineering, California Institute of Technology, Pasadena, California, 91125, USA*
[2]*Photonics Research Center, U.S. Military Academy, West Point, New York, 10996, USA*
[3]*Photonics Institute, Vienna University of Technology, 1040 Vienna, Austria*
[4]*Department of Physics, Norwegian University of Science and Technology, N-7491 Trondheim, Norway*
[5]*BAE Systems, P. O. Box 868, MER15-1813, Nashua, New Hampshire, 03061-0868, USA*
[†] *These authors contributed equally to this work.*
[*]*marandi@caltech.edu*

## 1. Simulation

Here, we describe the methodology behind the simulations used for numerical modeling of our experiment. Using the notation of [1], the coupled wave equations describing the phase-matched nonlinear interaction of pump and signal in the crystal are given by

$$d_z A_\omega(z,t) = \kappa A_{2\omega} A_\omega^* - \frac{\alpha_\omega}{2} + \widehat{\mathcal{D}}_\omega A_\omega \tag{S1a}$$

$$d_z A_{2\omega}(z,t) = -\kappa A_\omega^2 - \frac{\alpha_{2\omega}}{2} - u\frac{\partial A_{2\omega}}{\partial t} + \widehat{\mathcal{D}}_{2\omega} A_{2\omega} \tag{S1b}$$

where the time coordinate is taken to be comoving with the group velocity of the signal wave. Additionally, the pump envelope phase has been shifted by $\pi/2$ to ensure real solutions, neglecting higher order dispersion. Here, subscripts ω and 2ω refer to the signal and pump, respectively. $A_j$, $j \in \{\omega, 2\omega\}$ describes the field envelope and is normalized such that $|A_j|^2$ gives the instantaneous power. κ is the nonlinear coupling coefficient which governs the strength of the nonlinear interaction, given by $\sqrt{2\eta_0}\omega d_{eff}/(w_0 n_\omega \sqrt{\pi n_{2\omega}} c)$ where $\eta_0$ is the impedance of free space, $d_{eff}$ is the effective nonlinearity, $n_j$ is the refractive index, and $w_0$ is the Gaussian beam waist in the crystal. $\alpha_j$ is the absorption coefficient, given by the material loss in the crystal. $u$ is the walk-off parameter between pump and signal. Finally, $\widehat{\mathcal{D}}_j = \sum_{m=2}^{\infty} \left[\frac{(-i)^{m+1} \beta_\omega^{(m)}}{m!}\right] \partial_t^m$ is the dispersion operator.

Simulations of the field envelope evolutions in the crystal are performed using the split-step Fourier method, in which the OPA process in the crystal is divided into fifty segments. In each segment, we solve the linear and nonlinear portions of the coupled wave equations as lumped elements. The nonlinear step is computed by solving the nonlinear terms in the coupled wave equations using the fourth-order Runge-Kutta method. This is followed by a linear filter containing the dispersion and loss for the crystal, which is applied in the frequency domain to the pump and signal. We calculate four orders of dispersion for both pump and signal from the Sellmeier equation given in [2].

The roundtrip propagation is modeled by a linear feedback loop which contains the frequency-dependent losses as well as the dispersion for all cavity elements. Specifically, for the $n^{th}$ roundtrip, the signal at the input of the crystal $A_\omega^{n+1}(0,t)$ is related to the output of the OPA process, $A_\omega^n(L,t)$ by the equation

$$A_\omega^{n+1}(0,t) = \mathcal{F}^{-1}\{e^{-\frac{\alpha(\Omega)}{2}} e^{-i\Phi(\Omega)} \mathcal{F}\{A_\omega^n(L,t)\}\}. \tag{S2}$$

Here, $\Omega$ is the normalized Fourier frequency and $\mathcal{F}$ and $\mathcal{F}^{-1}$ are the Fourier transform and inverse Fourier transform, respectively. $\alpha(\Omega)$ gives the round-trip loss of the signal, including the frequency-dependent losses from the output coupling, the AR coatings on the crystal interfaces, the cavity mirrors, and the residual atmospheric gases in the cavity after purging, modeled using data provided from the HITRAN database [3]. Similarly, $\Phi(\Omega) = \Delta T_{RT}(\pi c/\lambda_{2\omega} + \Omega) + \Delta\Phi(\Omega)$ gives the round-trip phase accumulated by the signal, measured relative to a perfectly synchronous signal pulse, as mentioned in the main text. $\Delta T_{RT}$ is the detuning and contributes to both a constant phase term, $\frac{\Delta T_{RT} \pi c}{\lambda_{2\omega}}$, and a linear phase term, $\Delta T_{RT}\Omega$, which accounts for the timing delay. Here, c is the speed of light and $\lambda_{2\omega}$ is the wavelength of the pump. $\Delta\Phi(\Omega)$ accounts for higher-order contributions to the phase due to the dispersion of the cavity mirrors, the AR coatings on the crystal, and the residual gas in the cavity.

**Reference**

1. M. Jankowski, A. Marandi, C. R. Phillips, R. Hamerly, K. A. Ingold, R. L. Byer, and M. M. Fejer, Phys. Rev. Lett. **120**, 053904 (2018).
2. J. Wei, J. M. Murray, J. O. Barnes, D. M. Krein, P. G. Schunemann, and S. Guha, Opt. Mater. Express, **8**, 485 (2018).
3. I. E. Gordon, L. S. Rothman, R. J. Hargreaves, R. Hashemi, E. V. Karlovets, F. M. Skinner, E. K. Conway, C. Hill, R. V. Kochanov, Y. Tan, P. Wcisło, A. A. Finenko, K. Nelson, P. F. Bernath, M. Birk, V. Boudon, A. Campargue, K. V. Chance, A. Coustenis, B. J. Drouin, J. –M. Flaud, R. R. Gamache, J. T. Hodges, D. Jacquemart, E. J. Mlawer, A. V. Nikitin, V. I. Perevalov, M. Rotger, J. Tennyson, G. C. Toon, H. Tran, V. G. Tyuterev, E. M. Adkins, A. Baker, A. Barbe, E. Canè, A. G. Császár, A. Dudaryonok, O. Egorov, A. J. Fleisher, H. Fleurbaey, A. Foltynowicz, T. Furtenbacher, J. J. Harrison, J. –M. Hartmann, V. –M. Horneman, X. Huang, T. Karman, J. Karns,

S. Kassi, I. Kleiner, V. Kofman, F. Kwabia–Tchana, N. N. Lavrentieva, T. J. Lee, D. A. Long, A. A. Lukashevskaya, O. M. Lyulin, V. Yu. Makhnev, W. Matt, S. T. Massie, M. Melosso, S. N. Mikhailenko, D. Mondelain, H. S. P. Müller, O. V. Naumenko, A. Perrin, O. L. Polyansky, E. Raddaoui, P. L. Raston, Z. D. Reed, M. Rey, C. Richard, R. Tóbiás, I. Sadiek, D. W. Schwenke, E. Starikova, K. Sung, F. Tamassia, S. A. Tashkun, J. Vander Auwera, I. A. Vasilenko, A. A. Vigasin, G. L. Villanueva, B. Vispoel, G. Wagner, A. Yachmenev, and S. N. Yurchenko, Journal of Quantitative Spectroscopy and Radiative Transfer **277**, 107949 (2022).